\documentclass[aps,preprint]{revtex4-2}
\usepackage{graphicx}
\usepackage{dcolumn}
\usepackage{bm}
\usepackage{color}
\usepackage{comment}
\usepackage{appendix}

\begin{document}
\title{Confinement-induced toughening in delignified wood-polyacrylamide composite hydrogels}
\author{Kohei Shimogata$^{1}$ and Tetsuo Yamaguchi$^{1}$}
\affiliation{$^1$Department of Biomaterial Sciences, The University of Tokyo, Tokyo 113-8657 Japan}
\date{\today}

\begin{abstract}
With growing interest in environmental conservation, polymer composites incorporating wood-derived materials have attracted significant attention. 
While research has focused extensively on hard composites utilizing delignified wood or cellulose nanofibers (CNF) and soft composites reinforced with CNF, there have been a limited number of studies on the development of soft composites that utilize the porous cellular structure of wood, with few exceptions. 
 In this study, we fabricated soft composites by infiltrating delignified wood with polyacrylamide hydrogel and subjected them to uniaxial compression tests. The results demonstrated that these materials exhibited far superior strength and toughness compared to either delignified wood or polyacrylamide hydrogel alone. Through systematic investigation and theoretical analysis, we identified the mechanisms underlying these enhancements in strength and toughness. Specifically, in addition to the sacrificial bond breakage known to enhance toughness in double-network hydrogels, we propose that mechanical confinement, arising from the soft, incompressible hydrogel filling the hollow cell lumens, generates excess stress and acts to resist compressive deformation.
This approach, which utilizes the cellular structure as a filler, suggests the potential for creating a new class of high-toughness composite materials.
\end{abstract}

\maketitle

\section{Introduction}
In recent years, the growing demand for environmental conservation has created a need for plant-based materials to replace petroleum-derived ones\cite{Wool}. Wood is considered one of the most promising materials to meet this demand\cite{Berglund2018, Wang2021, Mao2022}. Wood is an assembly of hollow pipes enclosed by cell walls composed of three main constituents: an aligned cellulose skeleton, hemicellulose, and lignin\cite{Speck2011, Salmen2004, Jiang2018}. Its porosity, anisotropy, and the distribution of these three components give rise to its unique mechanical properties.
Recently, composite materials combining polymers with tissue structures of wood\cite{Berglund2018, Jiang2018, Zhu2023, Fu2018}, other than those with wood-derived cellulose nanofibers (CNF)\cite{Sharma2019, Fujisawa2021}, have been developed. In the polymer-wood composites, delignified wood\cite{Kumar2021, Horikawa2022}, which retains the hierarchical structure of wood and the aligned cellulose framework, is utilized as the filler. Since delignified wood becomes brittle when exposed to moisture, many composites incorporating it are designed to function as hydrophobic, hard materials\cite{Zhu2023, Fu2018}.

Meanwhile, research is also being conducted on soft composite materials derived from wood. In many of these materials, CNFs are combined with elastomers or hydrogels \cite{Trager2018, Lu2021, Chen2017, Chen2020}. Hydrogels are cross-linked polymers that swell upon absorbing water. Although they hold promise as functional materials, their compressive fracture strength is very low\cite{YamaguchiGels2018}; typically only a few hundred KPa at most\cite{Gong2010}. Over the past two decades, various types of hydrogels with enhanced strength and toughness have been developed. A representative example is the double-network (DN) hydrogel, which features an interpenetrating polymer network structure composed of a hard, brittle network and a soft, stretchable network\cite{Gong2010, Gong2003}. DN hydrogels exhibit exceptionally high strength and toughness compared to conventional hydrogels. Recently, it has been reported that composites made from delignified wood and hydrogel\cite{Wang2020}, which is considered one of the DN hydrogels, exhibit tensile strengths as high as 20 MPa\cite{Dong2022}. However, the mechanisms underlying this enhanced strength remain to be elucidated.

In this study, we fabricated delignified wood-polyacrylamide composite hydrogels by immersing aqueous solution of acrylamide monomers into delignified wood, and then by polymerizing and crosslinking the monomers. We conducted uniaxial compression tests and confirmed significant strengthening and toughening. As discussed later, we attribute the mechanisms to the mechanical confinement, as well as the sacrificial bond breakage.

\section{Experiments}
\subsection{Materials}
10 mm $\times$ 10 mm $\times$ 10 mm balsa wood blocks, glacial acetic acid ($\mathrm{CH_{3}COOH}$, Fuji Film Wako Pure Chemicals), sodium acetate trihydrate ($\mathrm{CH_{3}COONa\cdot 3H_{2}O}$, Sigma-Aldrich), and sodium chlorite ($\mathrm{NaClO_{2}}$, Sigma-Aldrich) were applied for the delignification of wood samples. Acrylamide monomer (Fuji Film Wako Pure Chemicals), N,N'-methylenebisacrylamide (BIS, Sigma-Aldrich) as the cross-linker, ammonium persulfate (APS, Sigma-Aldrich) as the polymerization initiator, and N,N,N'N'-tetramethylethylenediamine (TEMED, Sigma-Aldrich) as the polymerization accelerator were used for preparing polyacrylamide hydrogels. All chemicals were used as received. 

\subsection{Preparation of delignified wood}
Glacial acetic acid and sodium acetate trihydrate were added to water to make a buffer solution of 0.05 M acetic acid - sodium acetate, and a solution of 5 wt\% sodium chlorite was prepared with a pH of about $4\sim5$. The solution was heated to $80^{\circ}\mathrm{C}$, and balsa wood blocks reacted in the solution. After the reaction, the blocks were washed with deionized water and stored in ethanol. The delignified wood for the compression test was washed and stored in water.

\subsection{Preparation of delignified wood-polyacrylamide composite hydrogel}
First, acrylamide monomer solutions with cross-linker and 0.1 wt\% APS were prepared. Monomer concentrations were 10, 20, and 40 wt\%. Three cross-linker mass fractions of 0.005, 0.01, and 0.02 against the acrylamide monomer (i.e., hydrogel samples having three different molecular weights between adjacent cross-linkers) were prepared. 
In total, solutions with nine different combinations of monomer/cross-linker concentrations were prepared.

The delignified wood blocks were immersed in an acrylamide monomer solution, vacuumed for 30 minutes to remove trapped air inside the sample and to accelerate the immersion of the solution, and then stored in a refrigerator for 24 hours.

The delignified wood blocks were transferred from the acrylamide monomer solution to a 0.25 wt\% TEMED solution and stored for 5 minutes. After taking the sample out of the solution, the samples were sealed and then stored at room temperature for 24 hours. At least four samples were prepared for each condition. For comparison, polyacrylamide hydrogels with nine different monomer/cross-linker concentrations were also prepared in the same manner. Fig.1(a) shows the prepared composite hydrogel samples. 

\begin{figure}[htbp]
    \centering
    \includegraphics[keepaspectratio, width=\linewidth]{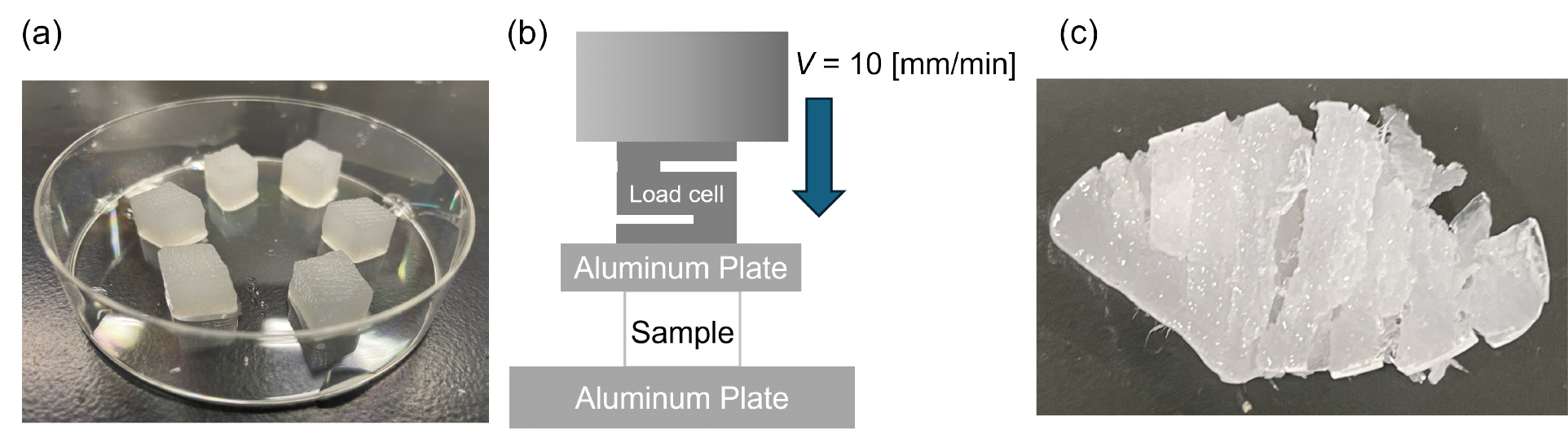}
     \caption{(a) Delignified wood-polyacrylamide composite hydrogels, (b) a schematic of the uniaxial compression test, and (c) a composite specimen after the compression test.}
     \label{fig1}
\end{figure}

\subsection{Uniaxial compression test}
Fig.1(b) shows a schematic of the experimental setup. Uniaxial compression tests were conducted on composite hydrogel, delignified wood, and polyacrylamide hydrogel alone, using a compression testing machine (SHIMADZU MST-X) at 10 mm/min. The direction of the compression of the wood was longitudinal (parallel to the fiber direction). For both the composite hydrogels and the polyacrylamide hydrogels, the samples were maintained as prepared.
Due to the compression, composite hydrogels are fractured into several fractions, as shown in Fig.1(c). 

\subsection{Scanning Electron Microscope (SEM) observation}
The surfaces of untreated wood, delignified wood, and composite hydrogel were observed using a scanning electron microscope (Hitachi High-Technologies S-4800). The applied voltage was 1.0 KV and the magnifications were 1,000$\times$, 2,500$\times$, 10,000$\times$, 25,000$\times$, and 80,000$\times$, respectively.

\subsection{Nitrogen absorption test}
Balsa wood powder and delignified wood powder were prepared. The composite hydrogels were air-dried, and the powder was obtained by scraping with a metal file. The pretreatment was carried out by vacuum heating at $80^{\circ}\mathrm{C}$ for one day and night\cite{Yang2019}. Nitrogen was then adsorbed at 77 K using a specific surface area/pore size analyzer (Microtrac Bell Belsorp mini II), and the ratio of adsorption equilibrium pressure to saturation vapor pressure, $p/p(0)$, and adsorption volume, $\Omega$, were measured.

\section{Results and Discussion}
\subsection{Compression experiments}
Fig.2 shows representative stress-strain curves 
obtained from uniaxial compression tests for delignified wood alone, polyacrylamide hydrogel alone, and composite hydrogel. The concentration of the monomer solution was 20 wt\%, and the cross-linker mass fraction 
in both composite hydrogel and polyacrylamide hydrogel
was 0.02. 

As clearly seen, the overall stress level of the composite hydrogel is much higher than that of the other two types of samples. 

\begin{figure}[htbp]
    \centering
    \includegraphics[keepaspectratio, width= 0.6 \linewidth]{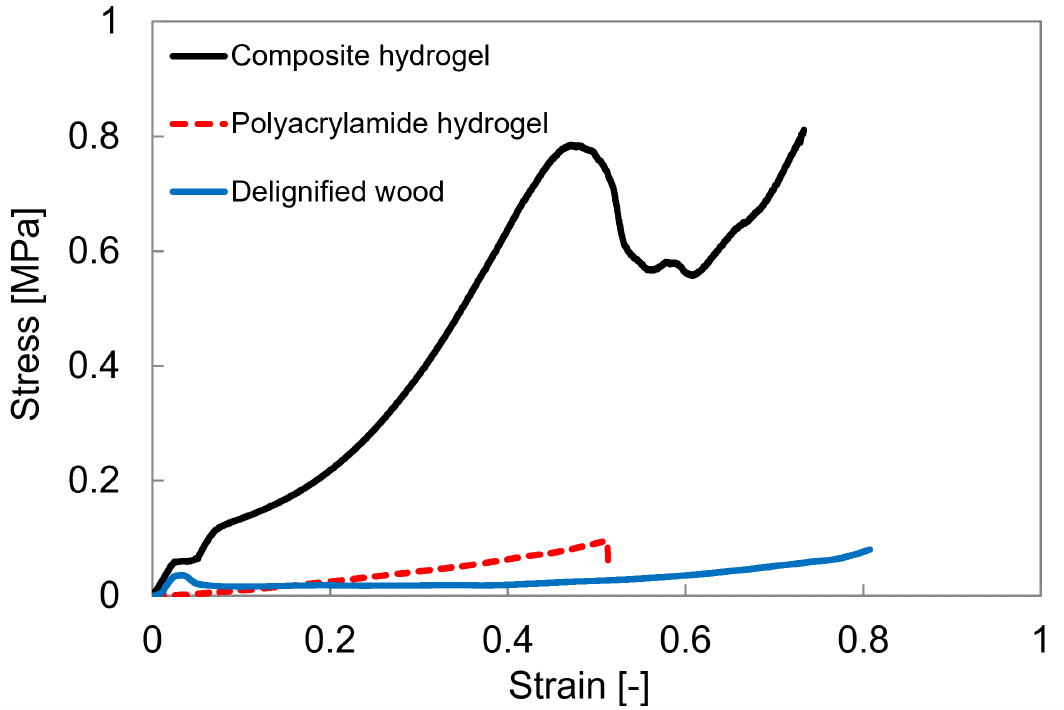}
    \caption{Representative stress-strain curves for the composite hydrogel, delignified wood alone, and polyacrylamide gel alone.}
    \label{fig2}
\end{figure}

A detailed examination of the measured stress-strain curves reveals several interesting features. In the low-strain region ($\epsilon \sim$ 0.03), where a change in the slope of the stress-strain curve indicates initial failure of the specimen, the behavior of this composite hydrogel resembles that of delignified wood. This suggests that the behavior at the early stage is governed by the delignified wood matrix within the composite hydrogel. In contrast, the strain at which macroscopic failure occurs ($\epsilon \sim$ 0.5) is comparable to that of polyacrylamide hydrogels. A key point here is that, unlike in tensile testing, the fractured fragments remain in place (see Fig.1(c)) during compression testing; consequently, the stress does not necessarily drop to zero beyond this point.

Fig.3 summarizes the stress-strain curves for all types of samples. Except in cases where both the monomer concentration ($C_{M}$) and the cross-linker mass fraction ($\phi_{CL}$) were low (see Fig.3(a)), the compressive stress of the composite hydrogels was significantly higher than that of the polyacrylamide hydrogels or the delignified wood samples.
Furthermore, while the strain at final break remained nearly constant (approximately 0.5), the stress at final break tended to increase with higher monomer concentrations.

\begin{figure}[htbp]
    \centering
    \includegraphics[keepaspectratio, width=\linewidth]{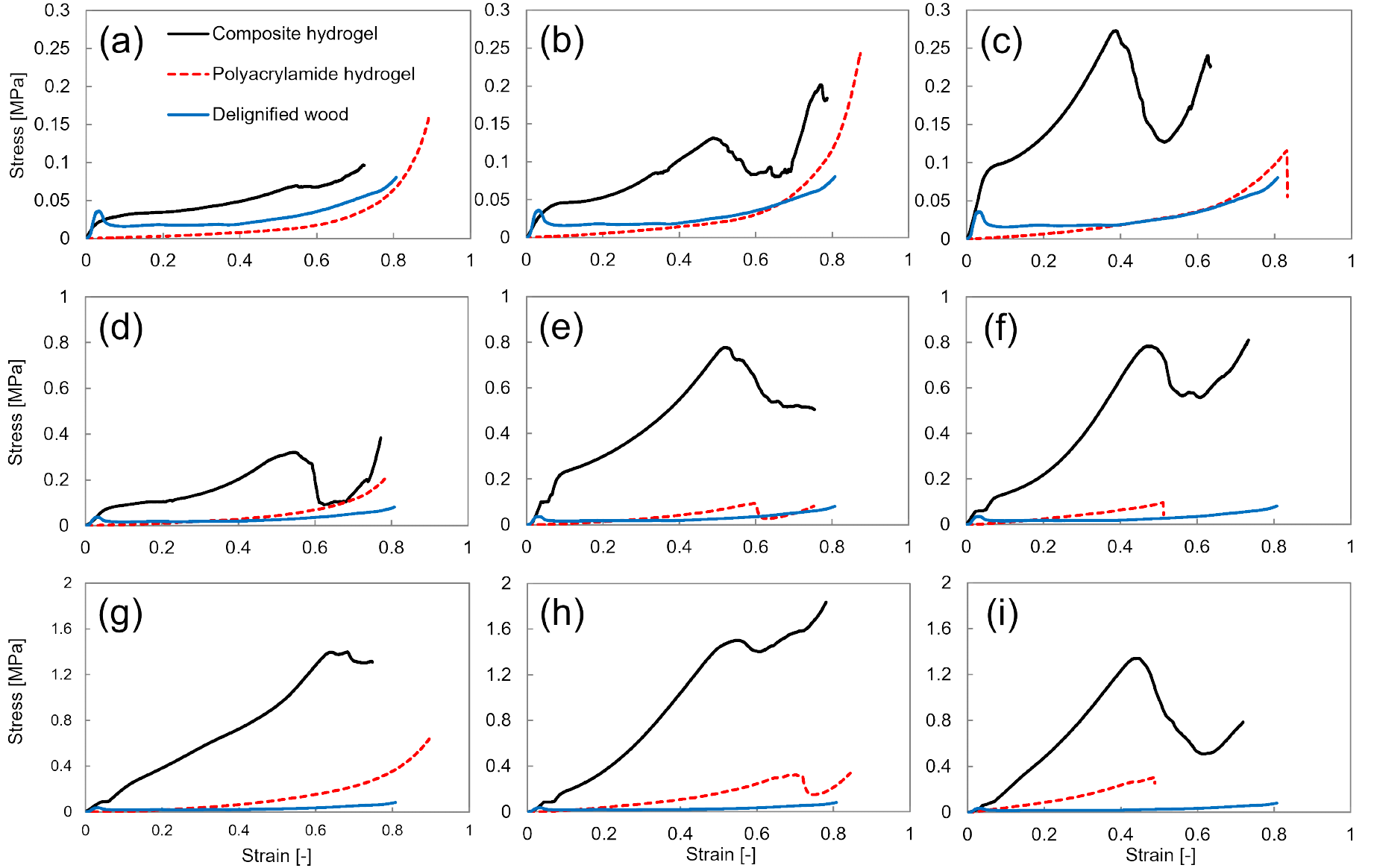}
    \caption{Compressive stress-strain curves for composite hydrogels, polyacrylamide hydrogels, and delignified wood. Monomer concentrations $C_{M}$ were (a)(b)(c) 10 wt\%, (d)(e)(f) 20 wt\%, and (g)(h)(i) 40 wt\%, and cross-linker mass fractions $\phi_{CL}$ were (a)(d)(g) 0.005, (b)(e)(h) 0.01, and (c)(f)(i) 0.02 against the acrylamide monomer.}
    \label{fig3}
\end{figure}

Next, we discuss the dependence of the cross-linker concentration $C_{CL} = C_{M} \phi_{CL}$ on various mechanical parameters, defined in Fig.4(a). Fig.4(b) shows the relationship between cross-linker concentration and Young's modulus $E$ for both the composite hydrogel and the polyacrylamide hydrogel. The Young's modulus of delignified wood is also plotted for comparison. The three distinct curves correspond to three different monomer concentrations. It is evident that the Young's modulus of the composite hydrogel is significantly higher than that of the polyacrylamide hydrogel. In addition, a strong dependence of the cross-linker concentration and a non-monotonic variation in Young's modulus are observed. Furthermore, there is a marked deviation from the composite law, which represents the weighted average of the Young's moduli of the two constituents. Fig.4(c) plots the stress at initial failure, $\sigma_{ini}$, against the cross-linker concentration. Despite some scatter, a trend similar to that of the Young's modulus, specifically, non-monotonic behavior, is evident.
As shown in Figs.4(d) and 4(e), the stress at final break, $\sigma_{f}$, and the fracture energy, $W$, exhibit a similar non-monotonic dependence.
In contrast, a different trend is observed for the strain at final break ($\epsilon_{f}$; see Fig.4(f)).
As previously mentioned, the value remains approximately 0.5, regardless of the cross-linker or monomer concentration.

\begin{figure}[htbp]
    \centering
    \includegraphics[keepaspectratio, width=\linewidth]{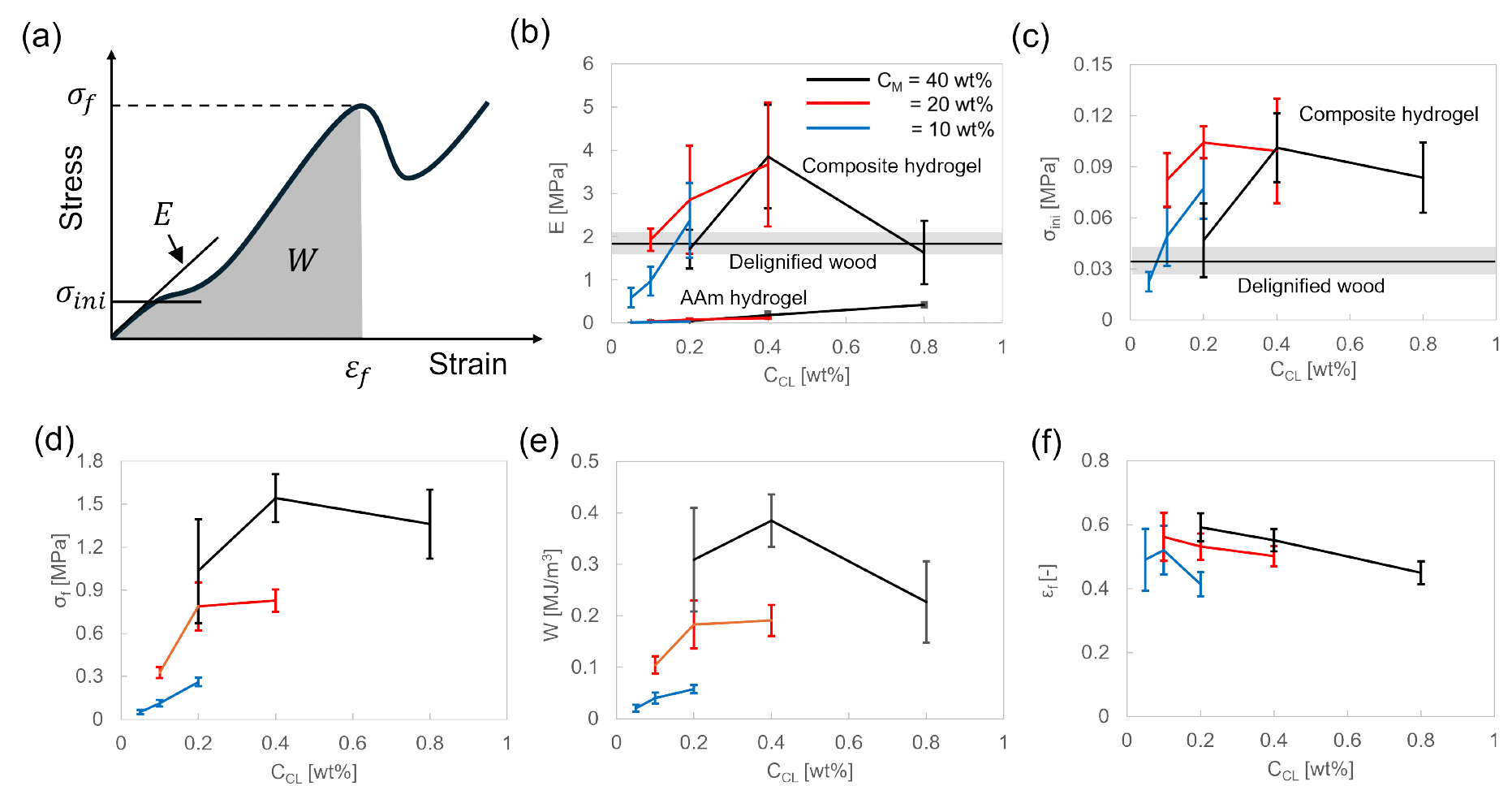}
    \caption{Summary of mechanical properties. (a) Definitions of mechanical parameters. Cross-linker concentration dependence on (b) Young's modulus, (c) Stress at initial failure, (d) Stress at final break, (e) Fracture energy, and (f) Strain at final break.}
    \label{fig4}
\end{figure}

\subsection{Stress enhancement mechanisms for composite hydrogels}
In the previous section, we confirmed that the compressive stress in the composite hydrogel was much larger than that generated in the delignified wood and polyacrylamide hydrogel. What causes this stress enhancement in the composite hydrogel?

\subsubsection{Structural analyses}
To elucidate the underlying mechanism, we first analyzed the microstructure of delignified wood and the composite hydrogel.

Fig.5(a) shows SEM images of the cross-section of the delignified wood. Cell lumens approximately 50 $\mu m$ in size are clearly visible. Closer inspection of the tissue structure reveals the presence of numerous mesopores. In contrast, the SEM image of the composite hydrogel shown in Fig.5(b) reveals that the cell lumens are closed (although not clearly visible in this image, this can be confirmed via optical microscopy). Furthermore, no mesopores are observed. This suggests that these pores are filled with polyacrylamide hydrogel.

To further investigate the mesopores, we performed nitrogen adsorption tests and analyzed the data\cite{Kaneko1994}. Fig.5(c) shows the size distributions of mesopores for unmodified wood, delignified wood, and composite hydrogel. As can be clearly seen, the delignified wood has a large number of mesopores ranging from 2 to 40 nm in diameter in the cell walls and has a distinct peak around 5 nm. It has been pointed out that these mesopores are formed due to the delignification process, where lignin is removed from the cellulose matrix. The pre-existing lignin domains are not isolated but are connected with other domains or the cell lumen. As will be discussed later, this interconnected void network is filled with polyacrylamide hydrogel and forms an interpenetrated-network structure, which provides a strong mechanical interlocking of polyacrylamide hydrogel onto the delignified wood substrate and significant toughening in composite hydrogels.

\begin{figure}[hbtp]
    \centering
    \includegraphics[keepaspectratio, width=\linewidth]{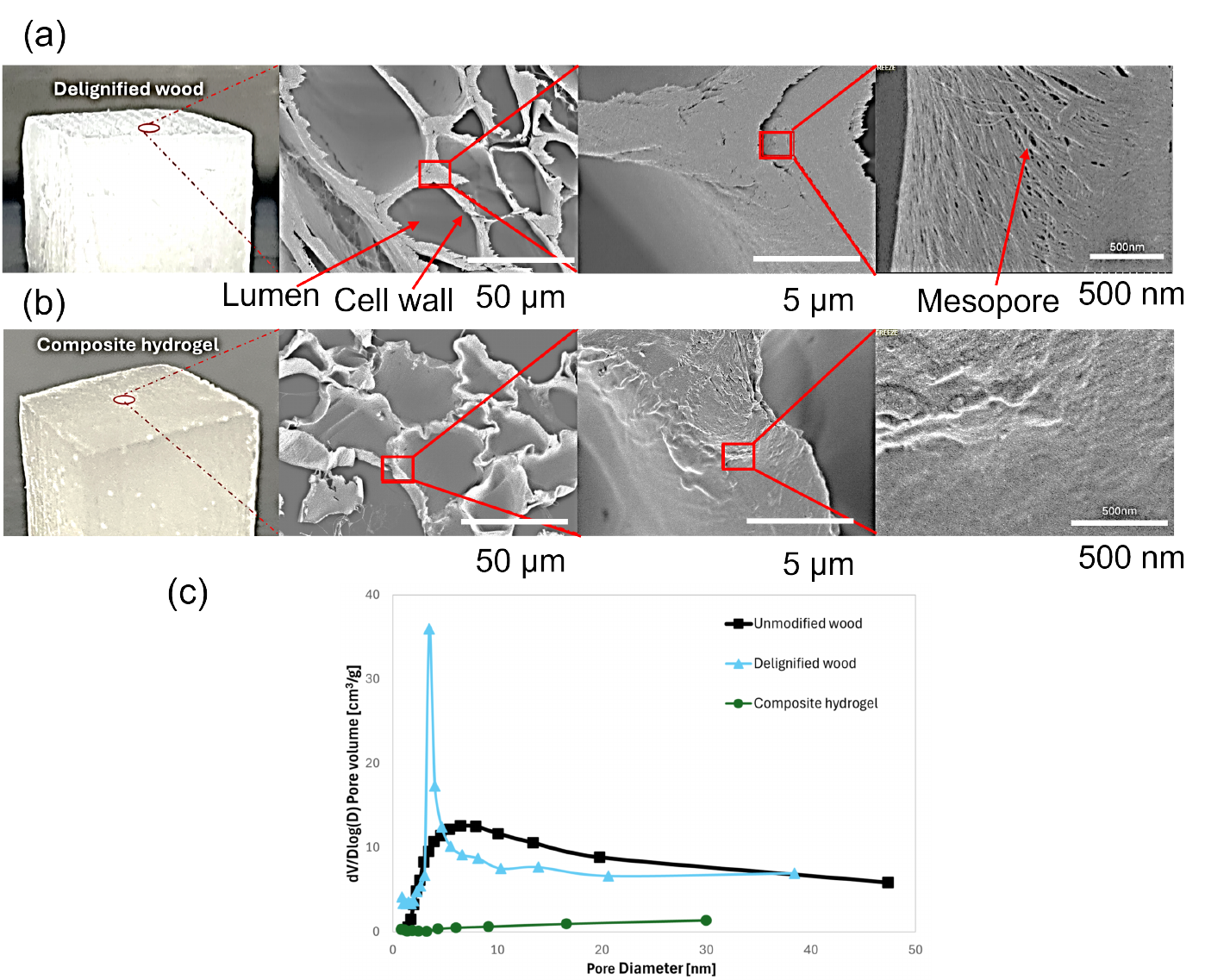}
    \caption{Optical Microscope and SEM images for (a) delignified wood and (b) composite hydrogel. (c) Pore size distributions of unmodified wood, delignified wood, and composite hydrogel.}
    \label{fig5}
\end{figure}

\subsubsection{Modeling of elastic deformations and initial failure for composite hydrogels}
Based on mechanical measurements and structural analysis, we constructed a theoretical model to calculate the Young's modulus and the stress at initial failure for the composite hydrogel.  

Fig.6(a) shows a schematic diagram of the cell wall structure of delignified wood. The long axis of the cells is aligned parallel to the direction of compression. Figs.6(b) and 6(c) are the schematic side views of the delignified wood and the composite hydrogel, respectively. Let the cell length be $l_0$, and the lumen radius be $r_0$. Here, we consider the case where a vertical compressive displacement $\delta$ is applied to a single cell. When the axial compression is applied, the cell wall starts to deform, generating elastic stress.  
Meanwhile, compression of the composite hydrogel also induces compression of the internal hydrogel cylinder.

\begin{figure}[hbtp]
    \centering
    \includegraphics[keepaspectratio, width=0.8\linewidth]{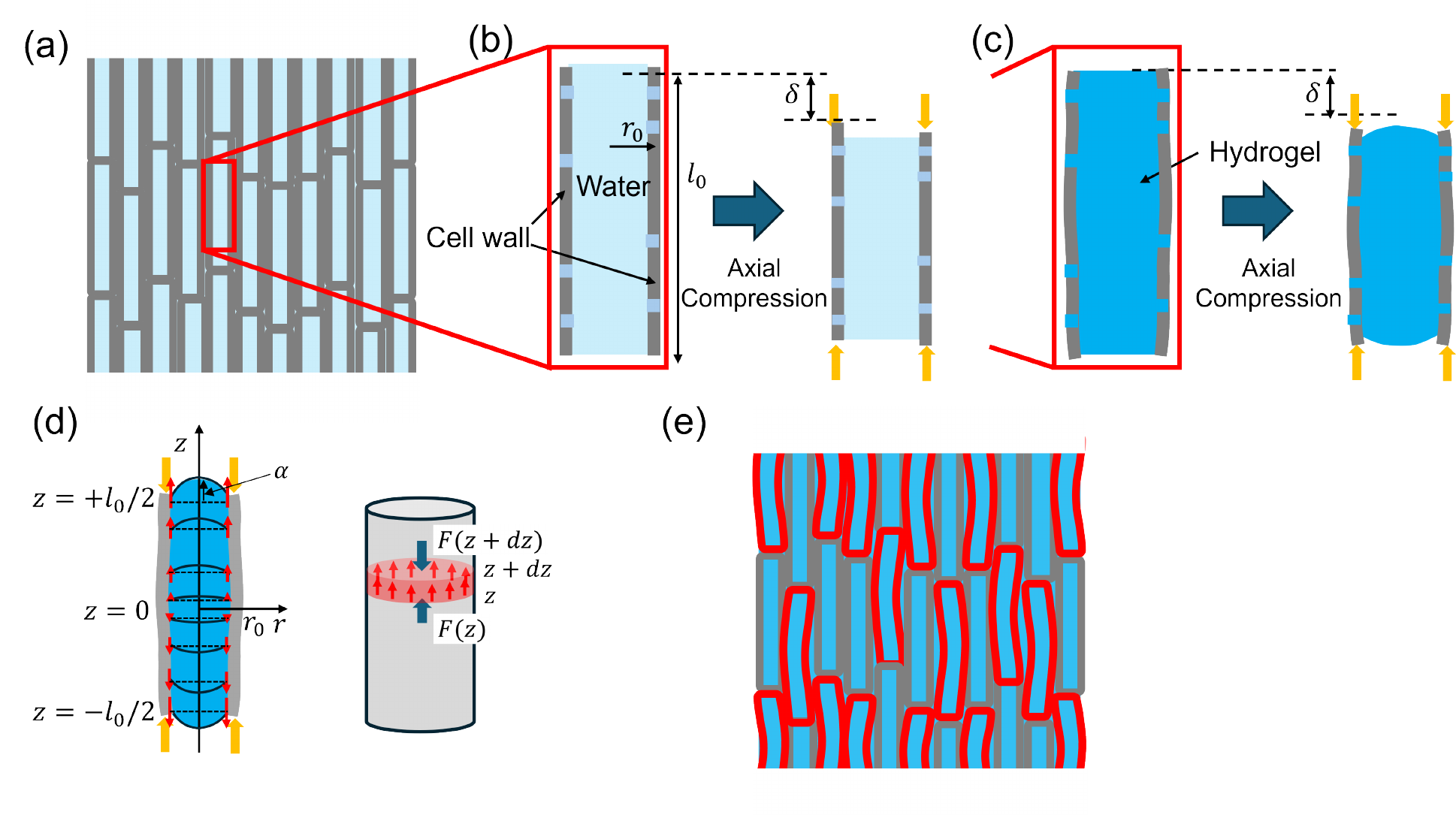}
    \caption{(a) Schematic diagram of the cell wall structure of delignified wood. Schematic side views of (b) delignified wood and (c) composite hydrogel. (d) Deformation profile of a single cell filled with hydrogel. (e) Local buckling of hydrogel-filled-cells at initial failure.}
    \label{fig6}
\end{figure}

If the cell wall is easily compressed while its lumen radius is kept constant due to the limited stretchability of the cell wall along the circumferential (lateral) direction, the hydrogel cannot deform uniformly; 
instead, to maintain a constant volume, it is forced to bulge out in a Poiseuille-like manner.

Fig.6(d) shows the deformation profile during axial compression. Assuming that the deformation is symmetric in the axial and vertical directions and that the hydrogel is incompressible, the vertical displacement can be expressed by the following equation.

\begin{equation}
    C(r, z) = - \delta \frac{z}{l_0} + \alpha \{1 - \Big( \frac{r}{r_0} \Big)^2 \} \frac{2z}{l_0},
    \label{eq1}
\end{equation}
where $\alpha$ is the protrusion displacement on the top surface ($z = l_0/2$) of the hydrogel. From volume conservation, we obtain the following relation. 

\begin{equation}
    \pi r_0^2 l_0 = \pi r_0^2 (l_0 - \delta) + 2 \times \frac{\pi}{2} r_0^2 \alpha, 
    \label{eq2}
\end{equation}
which results in $\alpha = \delta$. 
By applying Eqs.(1) and (2), we can calculate the shear stress acting on the cell wall at height $z$. 

\begin{equation}
    \tau(z) = - G \frac{\partial C}{\partial r}|_{r=r_0}  = 4G \frac{z}{r_0} \epsilon. 
    \label{eq3}
\end{equation}
where $G$ is the shear modulus of the inner hydrogel and $\epsilon = \delta/l_0$ is the applied compressive strain.

Then we calculate the force acting on the cell wall. By considering the force balance of the infinitesimal ring-like region from $z$ to $z+dz$, we obtain the following equation. 

\begin{equation}
  F(z+dz) - F(z) = 2 \pi r_0 \tau (z) dz,
  \label{eq4}
\end{equation}
where $F(z)$ is the compressive force acting on the cell wall at height $z$. 

Applying Eq.(4) and the mechanical condition $F(0) = \pi r_0^2 E_{CW} \epsilon$ (restoring force of the cell wall itself, where $E_{CW}$ is the nominal Young's modulus) at the center, we obtain the compressive force acting on the top surface of the cell wall. 

\begin{equation}
  F(l_0/2) = \pi ( E_{CW} r_0^2 + G l_{0}^{2}) \epsilon.
  \label{eq5}
\end{equation}
 
Finally, we obtain a formula for the Young's modulus, 

\begin{equation}
 E = \frac{\sigma}{\epsilon} = \frac{F(l_0/2)}{\pi r_0^2 \epsilon} = E_{CW} + G \Big(\frac{l_0} {r_0} \Big)^2.
 \label{eq6}
\end{equation}

Let us compare the theoretical result (Eq.(6)) with experimental data. 

First, for delignified wood samples, a cell lumen is filled with water (liquid) and no elastic stress is generated (i.e., $E_{CW} = E_{DW}$ and $G = 0$), 
thus the corresponding Young's modulus is just given by $E = E_{DW}$ ($=$ 1.83 MPa, the thin black line in Fig. 4(b)). 

Then, in the case of the composite hydrogel, the modulus should be given by Eq.(6). 
However, as shown in Fig.4(b), almost no contribution of the cell wall to elasticity is observed ($E_{CW} \approx 0$); the elasticity appears to be determined solely by the polyacrylamide hydrogel. 

\begin{equation}
E_{comp} \approx \frac{E_{AAm}}{3} \Big(\frac{l_0}{r_0} \Big)^2, 
\end{equation}
where $E_{AAm}$ is the Young's modulus for polyacrylamide hydrogel. 

This is likely because out-of-plane or wavy deformation occurs on the cell wall during the curing process, resulting in significant bending deformation upon compression and the loss of the cell wall's restoring force. It would be interesting to construct an ideal model system featuring straight, vertically oriented cell walls filled with hydrogel to demonstrate that the Young's modulus of the composite hydrogel can be expressed as the sum of the cell wall's restoring force and the shear force generated by the hydrogel, as indicated by Eq.(6).

To validate the proposed model, we compared the theoretical results (Eq.(7)) with the experimental data and plotted the results in Fig.7(a). Upon examining the fitting, we confirmed good agreement when the slenderness ratio $l_0/r_0 = 9.1$. This ratio is in qualitative agreement with previous observations ($ \approx 40$)\cite{Galos2022}, but differs by a factor of four. The reason for the discrepancy is presumably due to the presence of bordered pits\cite{Gorges2026}, that are located on the cell walls and breaks the sealing, leading to the decrease in effective $l_0/r_0$ ratios. 
Note that the data point for $C_{CL} = 0.8 \text{ wt\%}$ in Fig.4(b) was excluded as an outlier. Although the reason for the significant deviation from linearity at this point is not entirely clear, possible causes include the extremely slow diffusion of the polymerization initiator (APS) within the confined space (i.e., the cell lumen) or an incomplete cross-linking reaction.

This mechanical confinement is generated only when a soft elastic material is confined inside a narrow space enclosed with a hard material or between hard materials\cite{Yamaguchi2018}. It is interesting to see that a large pre-factor $(l_0/r_0)^2$ appears commonly in various types of phenomena, such as adhesion\cite{Shull2004, Yamaguchi2006, Yamaguchi2006-2, Doi2007, Yamaguchi2012}, friction\cite{Kaneko2008, Yamaguchi2025}, and seismic damping\cite{Kelly2011}.

\begin{figure}[hbtp]
    \centering
    \includegraphics[keepaspectratio, width=0.8 \linewidth]{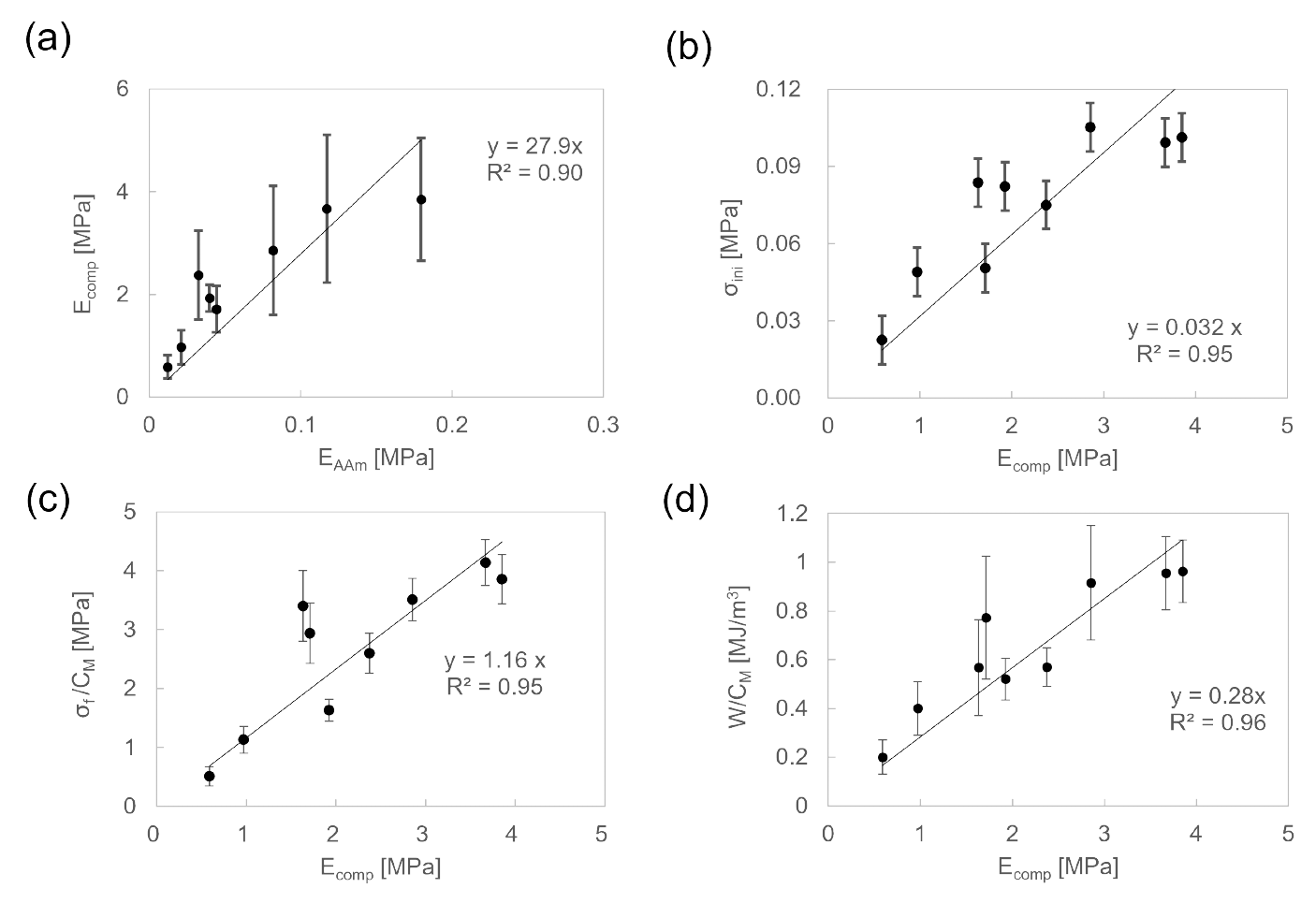}
    \caption{Responses of four mechanical parameters.  (a) $E_{comp}$ against $E_{AAm}$, (b) $\sigma_{i}$ against $E_{comp}$, (c) $\sigma_f/C_M$ against $E_{comp}$, and (d) $W/C_M$ against $E_{comp}$.}
    \label{fig7}
\end{figure}

Next, we consider the stress at initial failure. Here, we assume that the system exhibits "local" buckling of the cell walls at this stage, as shown schematically in Fig.7(e).
In other words, there is a heterogeneity in bending stiffness at the cell scale; some cells begin to buckle preferentially, while the remaining cells continue to undergo vertical compression. Furthermore, the buckling occurs in a single, isolated cell without constraint from neighboring (left or right) cells. According to Euler's buckling theory, the critical strain $\gamma_c$ for buckling is determined by the following formula:

\begin{equation}
    \gamma_c = \frac{\pi^2 E_{AAm} I_{CW}}{l_0^2} \frac{1}{\pi r_0^2} \frac{1}{E_{AAm}}  = \frac{\pi^2}{4} \Big( \frac{r_0}{l_0} \Big)^2
\end{equation}
where $I_{CW} = \pi r_0^4/4$ is the second moment of inertia for a single cell filled with hydrogel. Note here that the cell wall is not considered to contribute to the bending deformation because of the wavy deformation and negligible elasticity of the hydrogel-filling cell wall, as discussed in the earlier section. 

Finally, the stress at initial failure is given as;  
\begin{equation}
    \sigma_{ini} = E_{comp} \gamma_{c}.
\end{equation}  

In Fig.7(b), we plotted the relationship between $\sigma_{ini}$ and $E_{comp}$. We found a linear relation $\sigma_{ini} \approx 0.032 E_{comp}$ by fitting, which results in $l_0/r_0 =$ 9.9 with Eq.(8). This is in good agreement with the slenderness ratio obtained in fitting of the Young's modulus ($l_0/r_0 =$ 9.1). 

From these experimental observations and theoretical analysis, we inferred that the inner hydrogel starts to protrude and generate large stress during compression, and also that the initial failure is caused by the local buckling of cells filled with hydrogel.  

\subsubsection{Late stage behavior of composite hydrogels}
In the final part, we discuss the late-stage behavior and the toughening mechanisms of composite hydrogels. As discussed in the previous section, the cell lumens are filled with polyacrylamide hydrogels. 
When the composite hydrogel is further compressed, 
the protruded hydrogel touches the surrounding cell walls on the top and bottom, and then starts to generate excess hydrostatic pressure. This contributes to continuous increase in compression stress, until the late stage of the compression. Since this compression process is purely a geometrical one, the strain at final break is independent of the type of the sample.  
This mechanism should be verified with 
in-situ and detailed observations of the composite hydrogel. In addition, theoretical or numerical analyses are needed. This will be an important topic to be studied in the near future.

Instead, here we examined better collapse of the measured data. As shown in Figs.7(c) and 7(d), we confirmed the collapse in the curves by dividing the data by $C_{M}$. This indicates that, 

\begin{equation}
\sigma_{f} \approx \alpha_f C_M E_{comp}
\end{equation}
and 
\begin{equation}
W \approx \alpha_W C_M E_{comp}, 
\end{equation}
where $\alpha_f$ and $\alpha_W$ are numerical constants. 
It is not surprising to see that the similar behavior for $\sigma_f$ and $W$ is observed; this is due to the fact that the strain at final break $\epsilon_f$ is almost constant (see Fig.4(f)) and the shapes of the stress-strain curves are similar at least until the final break (Fig.3). On the other hand, the reasons why the monomer concentration $C_M$ affects the compressive stress at the late stage, unlike the cases for the Young's modulus or the stress at initial failure, are not so clear at this moment. This might be related to the fact that the composite hydrogel samples were tested not in equilibrium swelling but in as-prepared state; the partial chain length can change with the monomer concentration for the same cross-linker concentration. More detailed analysis or modeling of the fracture process is needed to understand the underlying phenomena. 

Though there is an unsolved problem mentioned above, the observed result is important for designing tough composite hydrogels; the Young's modulus induced by the confinement of the hydrogel inside a cell lumen, $E_{comp}$, controls the macroscopic mechanical properties at the late stage as well as at the early stage. This suggests that the more confinement we apply to the inner hydrogel (by increasing $l_0/r_0$ ratio), the higher strength and toughness a composite hydrogel exhibits. 
It would be interesting to create samples having larger $l_0/r_0$ ratios and to test whether the mechanical properties can be improved accordingly.

\subsubsection{Consideration for double network (DN) structure}
Finally, we discuss the connection between double-network (DN) hydrogel and our composite hydrogel. DN hydrogels is well-known as a tough hydrogel that combines a hard and brittle first network with a soft and stretchable second network, and the toughening mechanism is thought to be the sacrificial bond breakage of the first network at relatively low strain, which promotes energy dissipation and prevents fracture of the material as a whole\cite{Gong2003}.

The composite hydrogels fabricated in this study have a structure that is similar to that of DN hydrogel in structure and property; delignified wood can be considered as the first network consisting of cellulose-based cell walls, and polyacrylamide hydrogel filling the cell lumen as the second network. An image of a composite hydrogel after a compression test is shown in Fig.1(c), where several cracks appear along the radial structure, suggesting that the sacrificial bond breakage is active before the main rupture of the sample. 

However, remarkable enhancement in compressive stress cannot be explained by the sacrificial bond breakage alone, because the stress level is completely different from that for the first network (delignified wood) and the second network (polyacrylamide hydrogel), as clearly seen in Figs.2 and 3. 
In fact, there have been several studies on the mechanical effects of the double-network; analog DN models\cite{Feng2016}\cite{King2022} and molecular dynamics simulations\cite{Higuchi2018} in uniaxial tensile tests have shown that the initial stress peak is controlled by that of the first network and the final failure stress is determined by that of the second network, although successive bond breakage of the first network occurs and contributes to the tensile stress until large strains. 

In this study, the large enhancement in stress is considered to occur via mechanical confinement, where a soft component is confined by a hard one. It would be interesting to discuss the confinement (presumably nano-confinement) in the double-network hydrogels to understand the mechanisms for the stress enhancement.  

\section{Conclusion}
In this study, we fabricated soft yet tough composite hydrogel by combining delignified wood with polyacrylamide hydrogel while preserving its cellular structure. Uniaxial compression tests confirmed that the mechanical properties of this composite hydrogel were vastly superior to those of delignified wood or polyacrylamide hydrogel alone. Based on the experimental results and theoretical analysis, we examined the mechanisms underlying the strengthening and toughening of the composite hydrogel, focusing on the mechanical confinement and the breakage of sacrificial bonds within a double-network structure.

These mechanisms suggest a potential to toughen polymer composite materials, such as hydrogels, rubbers, and plastics, within a unified framework of utilizing porous structures as templates. Future research will aim to advance this work by developing tough polymeric materials that leverage porous or cellular structures, as well as soft composites composed entirely of biomass-derived components. These studies will be reported somewhere as separate works in near future.

\section*{Author contributions}
K.S. and T.Y. conceptualized this research. K.S conducted experiments, and T.Y. did theoretical analyses. K.S. and T.Y. analyzed the data and prepared the manuscript.

\section*{Conflicts of interest}
The authors declare that they have no conflicts of interests.

\section*{Data availability}
The datasets generated during and/or analyzed during the current study are available in the article and its Electronic Supplementary Information (ESI).

\section*{Acknowledgments}
This research was supported by JSPS KAKENHI Grant Number JP24K00595.
The support was also provided under the partnership agreement between the University of Tokyo and Sumitomo Mitsui Financial Group. 
We acknowledge M. Ikeda, S. Ozawa, and D. Okumura for their important suggestions. We also thank H. Suzuki and T. Uemura for their experimental support.

\end{document}